\documentclass[	
						superscriptaddress,%
						twocolumn,%
						a4paper,
						showkeys,
					]{revtex4-1}

\usepackage[english]{babel}
\usepackage[T1]{fontenc}
\usepackage{upgreek}
\usepackage{textcomp} 				
\usepackage[dvips]{graphicx}		
\usepackage{color}					
\usepackage[nice]{nicefrac} 		
\usepackage{amssymb,amsmath}		
\usepackage{array}					
\usepackage{ulem}

\renewcommand{\emph}[1]{\textit{#1}}

\definecolor{vert}{rgb}{0.5,0.758,0.5}
\definecolor{bleufonce}{rgb}{0,0,0.516}
\definecolor{orange}{rgb}{1,0.516,0}

\begin{document}

\title{Theory of antiferroelectric phase transitions}
\date{\today}
\author{Pierre Tol\'edano}
\affiliation{Laboratoire de Physique des Syst\`emes Complexes, Universit\'e de Picardie, 80000 Amiens, France}
\author{Mael Guennou}
\email{mael.guennou@list.lu}
\affiliation{Materials Research and Technology Department, Luxembourg Institute of Science and Technology, 41 rue du Brill, L-4422 Belvaux, Luxembourg}

\begin{abstract}
At variance with structural ferroic phase transitions which give rise to macroscopic tensors coupled to macroscopic fields, criteria defining antiferroelectric (AFE) phase transitions are still under discussion due to the absence of specific symmetry properties characterizing their existence. They are recognized by the proximity of a ferroelectric (FE) phase induced under applied electric field, with a double hysteresis loop relating the induced polarization to the electric field and a typical anomaly of the dielectric permittivity. Here, we show that there exist indeed symmetry criteria defining AFE transitions. They relate the local symmetry of the polar crystallographic sites emerging at an AFE phase transition with the macroscopic symmetry of the AFE phase. The dielectric properties of AFE transitions are deduced from a Landau theoretical model in which ferroelectric and ferrielectric phases are shown to stabilize as the result of specific symmetry-allowed couplings of the AFE order- parameter with the field-induced polarization.
\end{abstract}


\maketitle

\section{Introduction}

The relation between antiferroelectricity and ferroelectricity is traditionally assumed to be analogous to the relation between antiferromagnetism and ferromagnetism. However, there is an essential difference between the symmetry properties governing spin and dipole orderings. In the paramagnetic state, time-reversal symmetry, which is not a symmetry operation in real space, exists everywhere and is lost at the transition to the antiferromagnetic or ferromagnetic states, only surviving in combination with part of the symmetry operations associated with the crystallographic space group of the atomic structure \cite{Birss1966}. By contrast, in the paraelectric (PA) phase the symmetry operations of the crystal space-group are localized in space. As such, a paraelectric "state" does not exist by itself in the same sense as the paramagnetic state, and all crystal structures displaying non-polar symmetries can be potentially antiferroelectric. 

Definitions of antiferroelectricity involve the microscopic picture of antiparallel dipoles, the existence of an electric field induced FE phase associated with a double hysteresis loop relating the electric polarization and the field, and a characteristic anomaly of the dielectric permittivity \cite{Rabe2013,Tagantsev2013,Hlinka2014}. A symmetry-based definition of AFE transitions is still lacking, in spite of previous attempts \cite{Scott1974,Roos1976} inspired from a displacive picture, and focused on a definition for atomic displacement patterns associated to an AFE "soft-mode" as the driving mechanism of the PA to AFE transition. Such a definition is not well suited for describing the many important cases of order-disorder antiferroelectrics that represent a majority of the well-established AFE transitions. Besides, they are not linked to specific physical properties, so that the definition is of little use. This led some authors to state that antiferroelectricity was an ill-defined and hardly useful notion \cite{Levanyuk1969}. 

In this work we propose a definition of AFE transitions which stems from definite symmetry conditions fulfilled exclusively in AFE materials (section~\ref{sec:symmetry}). The dielectric properties characterizing AFE transitions are deduced from a Landau model in which ferroelectric (FE) and ferrielectric (FI) phases are shown to stabilize under specific field-induced couplings of the AFE order-parameter with the electric polarization (section~\ref{sec:landau}). In section~\ref{sec:analogy}, the traditional analogy between antiferroelectricity and antiferromagnetism, the concept of local polarization assumed in our theoretical description, and the nature of the AFE order-parameter are discussed. Last, in section~\ref{sec:summary} our results are summarized and the difference of our theoretical approach with Kittel's model of antiferroelectricity \cite{Kittel1951} is underlined. 

\section{Symmetry-based definition of AFE transitions}
\label{sec:symmetry}

PA--AFE transitions are structural transitions between phases displaying non-polar space-groups, which exhibit under applied electric field a typical dielectric behavior. A Landau symmetry analysis of the order-parameters associated with the PA--AFE transitions reported experimentally shows that the transitions occur exclusively in materials \emph{which do not have a stable FE phase in their phase diagram at zero fields}. It indicates that their dielectric properties are purely field-induced effects. Since only a limited number of structural phase transitions between non-polar space-groups exhibit such dielectric properties, PA--AFE transitions should obey restrictive and specific conditions. These conditions have so far not been found from the macroscopic symmetries of the PA or AFE phases alone. The specificity of PA--AFE transitions has therefore to be searched in the microscopic features of their transition mechanism. Here we show that two symmetry-based conditions have to be fulfilled for a phase transition to qualify as PA--AFE. A first condition deals with the local changes occurring at the transition. 

\emph{Condition 1: At the PA--AFE transition, a set of crystallographic sites undergo a symmetry lowering that results in the emergence of polar sites and give rise to a local polarization.}

\begin{figure}[htbp]
\begin{center}
\includegraphics[width=0.48\textwidth]{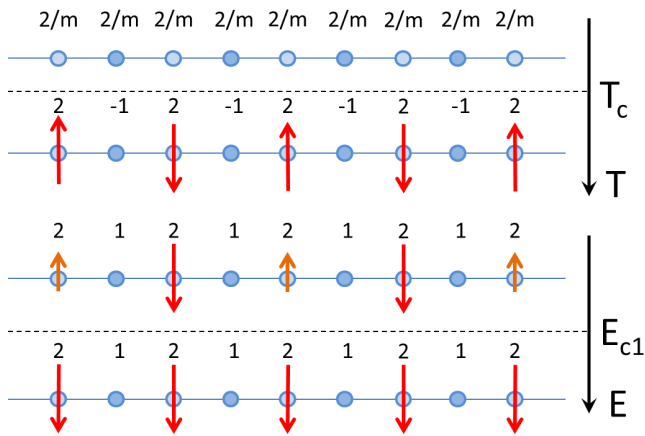}
\end{center}
\caption{Illustration of the local symmetry-breaking mechanism corresponding to Condition 1. In the PA phase, all sites of an atomic row have a non-polar site symmetry $2/m$. At the AFE transition at $T_c$ the symmetry of every other atomic site lowers to $2$. The sites located at mid distance preserve their inversion centre yielding the onset of antiparallel dipoles on the polar sites. An electric field oriented along the dipoles favours a ferrielectric configuration. Above a coercive field $E_{c1}$ a FE dipolar order arises.}
\label{fig:1D}
\end{figure}

Fig.~\ref{fig:1D} illustrates the local symmetry-breaking mechanism corresponding to condition 1 for a one-dimensional toy model. In the PA phase, all sites have the non-polar site symmetry $2/m$. At the PA--AFE transition, every other site acquires the polar site symmetry $2$ while the sites located in between keep their inversion center, which results in the onset of antiparallel local polarizations on the "active" polar sites, and a non-polar macroscopic symmetry. The principle is the same in real systems. Fig.~\ref{fig:2} (a)--(c) shows the onset of polar sites in the well-established examples of PA--AFE transitions reported in Cu(HCOO)$_2$.4H$_2$O \cite{Okada1965}, KCN \cite{Stokes1984} and PbZrO$_3$ \cite{Teslic1998}. They involve two or four sets of independent polar sites arising from non-polar sites at the transitions, which may carry anti-parallel arrays of dipoles.

The emergence of polar sites and local polarization at PA--AFE transitions constitutes a necessary condition, but the following additional condition is required for preserving the site symmetries at the macroscopic level, permitting a subsequent stabilization of a FE phase under applied electric field:

\begin{figure}[htbp]
\begin{center}
\includegraphics[width=0.4\textwidth]{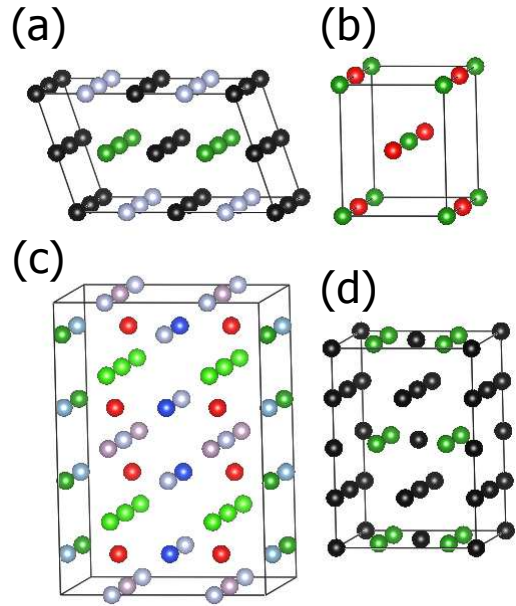}
\end{center}
\caption{Emerging polar sites at the AFE transitions in (a) Cu(HCOO)$_2$.4H$_2$O \cite{Omura2002}, (b) KCN \cite{Stokes1984}, (c) PbZrO$_3$ \cite{Teslic1998}, and (d) at the non-AFE transition in SrTiO$_3$. Coloured spheres indicate the polar sites carrying antiparallel dipoles in the AFE phase, whereby cristallographically equivalent sites are drawn in the same colour. At sites drawn in black, no local polarisation can emerge.}
\label{fig:2}
\end{figure}

\emph{Condition 2: The AFE space-group has a symmorphic polar subgroup coinciding with the local symmetry of emerging polar sites.}

This condition derives from the property that only symmetry operations of the AFE space-group forming a symmorphic group preserve the local site symmetries, while screw axes and glide planes modify the symmetry of the local sites. That the emergence of polar sites (condition 1) is not a sufficient condition for the existence of a PA--AFE transition can be exemplified by the ferroelastic transition in SrTiO$_3$, where the dielectric properties characterizing AFE materials have not been detected. Fig.~\ref{fig:2} (d) shows the onset of polar sites of symmetry $mm2$ at its cubic $Pm\overline 3m$ to tetragonal $I4/mcm$ transition. The symmorphic polar subgroups of $I4/mcm$ are $I4$, $C2$, $Cm$ and $P1$, the point-groups of which differ from $mm2$. Therefore, condition 2 is not fulfilled and the transition in SrTiO$_3$ does not have an AFE character, in spite of the arguable existence of a local polarization on the $mm2$ sites of its tetragonal phase. 

\begin{table*}[htbp]
\renewcommand{\arraystretch}{0.65}
\caption{Verification of conditions 1 and 2 for selected structural transitions in AFE and other materials listed in column (a). Other columns have the following meaning: (b) Space-group changes occurring at the transitions.  (c)  Crystallographic sites undergoing a lowering of their local symmetry to a polar point-group.  (d) Symmorphic polar subgroups of the low-symmetry phase space-groups coinciding with the emerging polar site symmetries. (e)  Couplings between the AFE transition order-parameter and the polarization allowing emergence of a field-induced polar phase. (f) Corresponding orientation(s) of the electric-field. (g) Space-group of the FE phase. In (e) and (f) except for CsH$_3$(SeO$_3$)$_2$ only couplings corresponding to  non-general directions of $P$ and $E$ are given. In (e), $\eta_1$ and $\eta_2$ are two components of the same transition order-parameter except for PbZrO$_3$ where they correspond to different order-parameters.}
\centering
\begin{tabular}{l >{$}l<{$} >{$}l<{$} >{$}l<{$} >{$}l<{$} >{$}l<{$} >{$}l<{$} }
\hline\hline
(a) & $(b)$ & $(c)$ & $(d)$ & $(e)$ & $(f)$ & $(g)$ \\
\hline
CsH$_3$(SeO$_3$)$_2$ 													& P\overline 1\rightarrow P\overline 1 	& 1a: \overline 1 \rightarrow 1 					& P1 		& \eta^2 P^2 & E  & P1\\
Cu(HCOO)$_2$.4H$_2$O 	  											& P2_1/a \rightarrow P2_1/a 						& 2a,2b,2c,2d: \overline 1\rightarrow 1  	& P1  	& \eta^2 P_z^2 & E_z & P2_1 \\                      
KCN 																					& Immm \rightarrow Pmmn 								& 2a, 2c: mmm \rightarrow mm2 						& Pmm2 	& \eta^2 P_z^2 & E_z & Imm2  \\
C$_4$O$_4$H$_2$ 		 													& I4/m \rightarrow P2_1/m 							& 2a,2b: 4/m \rightarrow m								& Pm,P1 & (\eta_1\eta_2,\eta_1^2-\eta_2^2)\times 	& E_{xy} & Cm  \\
																							& 																			& 4c: 2/m \rightarrow m 									&    		& (P_xP_y,P_x^2-P_y^2), 	 \\
																							& 																			& 4d: \overline 4 \rightarrow 1 					&    		& (\eta_1^2+\eta_2^2)P_z^2 & E_z & I4\\
NH$_4$H$_2$PO$_4$ 														& I\overline 42d \rightarrow P2_12_12_1 & 4a,4b: \overline 4 \rightarrow 1				& P1 		& \mathrm{idem} 	& E_{xy}, E_z & Cc, Fdd2 \\
DyVO$_4$  																		& I4_1/amd \rightarrow Imma 						& 4a, 4b: \overline 4m2 \rightarrow mm2		& Imm2,P1 & \eta^2 P_z^2 	& E_z  & I4_1md \\
																							& 																			& 16h: 2_{xy} \rightarrow 1								&         & \eta(P_x^2-P_y^2) & E_{xy} & Imm2 \\
BiVO$_4$																			& I4_1/a \rightarrow B2/b 							& 4a, 4b: \overline 4 \rightarrow 2 			& C2 &  \eta^2 P_z^2 & E_z & I4_1 \\
																							&							 													&																					&    & \eta (P_xP_y,P_x^2-P_y^2) & E_{xy} & Cc\\
TeO$_2$																				& P4_12_12 \rightarrow P2_12_12_1 			& 4a: 2\rightarrow 1 											& P1 & \eta^2 P_z^2 	& E_z & P4_1  \\
																							&																				& 8b: 1\rightarrow 1		 									& 				& \eta (P_x^2-P_y^2)  & E_{xy} & C2\\
PbZrO$_3$ 																		& Pm\overline 3m \rightarrow Pbam       & 3d: 4/mmm \rightarrow m,2 							& Pm, P2 & (\eta_1^2,\eta_2^2)\times P^2_{x,y,z}	& E_{x,y,z} & P4mm  \\
																							&																				& 1a, 1b: m\overline 3m \rightarrow m,1		& P1 	& (\eta_1^2,\eta_2^2)\times P^2_{xy,yz,zx}	& E_{xy,yz,zx} & Amm2 \\
NdP$_5$O$_{14}$  															& Pmna \rightarrow P2_1/b 							& 4e, 4f: 2 \rightarrow 1 								& P1 & \eta P_xP_y & E_{xy} & Pnc2 \\
																							&																				& 4h: m \rightarrow 1 										&    & \eta^2 P^2_z& E_z    & Pmn2_1\\
NH$_4$Cl 																			& Pm\overline 3m \rightarrow P\overline 43m &  \mathrm{none} \\
SrTiO$_3$ 			  														& Pm\overline 3m \rightarrow I4/mcm 		& 3d: 4/mmm \rightarrow mm2 							& I4, C2, Cm, P1 	& \mathrm{none} \\
\hline\hline
\end{tabular}
\label{tab}
\end{table*}

Fig.~\ref{fig:1D} shows the local effect of an applied electric field on the emerging polar sites: low fields oriented along the pre-existing local polarization favour a ferrielectric (FI) configuration, and above a coercitive $E_c$ field a FE dipolar order arises. The preservation of the polar site symmetry and local polarization by a symmorphic subgroup of the AFE space-group allows realizing at the macroscopic level a similar sequence of FI and FE phases induced from the AFE phase under suitably oriented fields. The space-group of the FE phase coincides with the symmetry of the PA phase under applied field; it may contain screw axes or glide planes and is always of higher symmetry than the symmorphic subgroup of the AFE phase. Depending on the orientation of the field, the FE phase may involve symmetry operations which do not belong to the AFE space group. The space-group of the FI phase is a common polar subgroup of the AFE and FE space-groups; it may also have a higher symmetry than the symmorphic space-group. 

A verification of conditions 1 and 2 for confirmed AFE materials is summarized in Table~\ref{tab}. It shows that the two conditions are unambiguously verified for the corresponding PA--AFE transitions, the highest polar site symmetry coinciding with the maximal symmorphic polar subgroup of the AFE space group. The table indicates the orientation of the electric fields stabilizing a FE phase. Also listed are a number of materials (BiVO$_4$, TeO$_2$, NdP$_5$O$_{14}$) which undergo transitions fulfilling conditions 1 and 2 but have not yet been recognized as antiferroelectric. It contains as well examples of transitions that do not verify condition 1 (NH$_4$Cl) or condition 2 (SrTiO$_3$). For some of the listed materials (DyVO$_4$, TeO$_2$, NdP$_5$O$_{14}$), the local polarization emerging at the transition results from a symmetry lowering on sites with an already polar site-symmetry in the PA phase, which illustrates the property that PA and AFE "states" do not exist per se but can only be defined in the context of a PA--AFE phase transition.

As will become apparent in the following, a variety of situations have to be taken into account in real systems. For instance, due to the high energy barrier that may exist between the FI and FE phases, the latter may not always be stabilized at high fields (sec.~\ref{sec:analogy}). Moreover, for specific couplings of the PA--AFE transition order-parameter with the polarization the FE phase becomes unstable, the field-induced FI-FE sequence of phases being replaced by a sequence of two isostructural FI phases (sec.~\ref{sec:landau}). Accordingly, the emergence of a FE phase is not a pre-requisite for PA--AFE transitions, and the realization of a double hysteresis loop under high fields can be effective or latent. Therefore, our theoretical analysis allows proposing the following symmetry-based definition which encompasses the variety of experimental situations:

\emph{PA--AFE transitions are structural transitions between non-polar phases where the symmetry of crystallographic polar sites emerging at the local scale coincides with the symmetry of a polar symmorphic subgroup of the AFE space-group, allowing the emergence of an electric field induced polar phase at the macroscopic scale.}

\section{Dielectric properties of AFE transitions}
\label{sec:landau}

Application of an electric field $E$ to a non-polar AFE phase induces a coupling between the AFE order-parameter, here labelled $\eta$, and the polarization $P$. The lowest degree coupling between $\eta$ and $P$ determines the stability and symmetry of a field-induced FE phase, and the orientation of the electric dipoles in the AFE phase. It also establishes the link with the dielectric anomalies typifying AFE transitions. Only couplings of the $\eta^2P^2$ or $\eta P^n$ ($n\ge 2$) types can be associated with an AFE transition, since only such couplings reflect the remarkable property of AFE transitions to occur in materials \emph{which do not have a stable polar phase at zero fields}. For a biquadratic $\eta^2P^2$ coupling, the dielectric properties of AFE transitions derive from the Landau potential: 
\begin{equation}
\phi(\eta,P,T) = \phi_0(T) +  \frac{\alpha}{2}\eta^2 + \frac{\beta}{4}\eta^4 +\frac{\gamma}{6}\eta^6 + \frac{P^2}{2\chi_0} + \frac{\delta}{2}\eta^2P^2 - EP 
\label{eq:potential}
\end{equation}
where $\alpha = a(T-Tc)$, and the other phenomenological coefficients are constant. This order-parameter expansion differs from the seminal model by Kittel \cite{Kittel1951} as it involves a single symmetry-breaking AFE order-parameter $\eta$, the polarization $P$ being a field-induced order-parameter. It expresses the property that a polar phase requires the mediation of the AFE order parameter to be stabilized upon application of an electric field. 

Minimizing $\phi$ with respect to $\eta$ and $P$ yields the equations of state:
\begin{eqnarray}
\eta(\alpha + \beta \eta^2 + \gamma \eta^4 + \delta P^2) & = & 0 \label{eq:eos1}\\
P(1+\delta\chi_0\eta^2) & = & \chi_0 E \label{eq:eos2}
\end{eqnarray}

At $E=0$, Eqs.~(\ref{eq:eos1}) and (\ref{eq:eos2}) yield two possible stable phases: the PA phase ($\eta=0$, $P=0$) and the AFE phase ($\eta\neq 0$, $P=0$). For $E\neq 0$ two phases are stabilized: a FE phase ($\eta =0$, $P\neq 0$) in which the AFE antiparallel dipole configuration is absent, and a phase in which the AFE ordering  ($\eta\neq 0$) has a non-zero total polarization ($P\neq 0$), i.e. having either a ferrielectric (FI) dipolar order or a "weak" ferroelectric (WF) order with a canting between antiparallel arrays of dipoles. 

\begin{figure}[htbp]
\begin{center}
\includegraphics[width=0.48\textwidth]{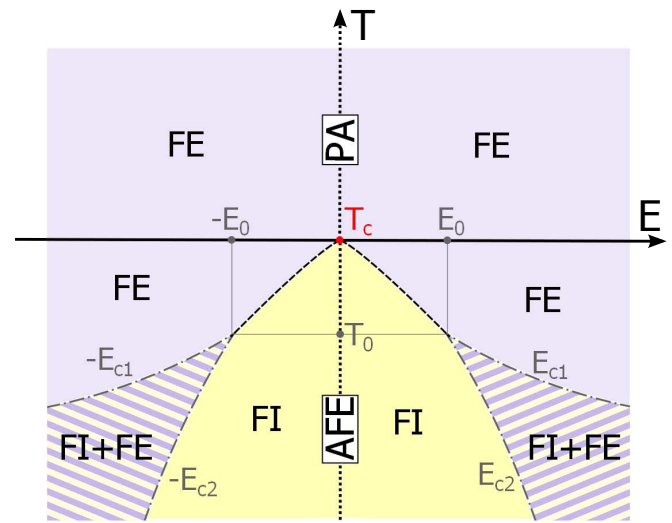}
\end{center}
\caption{Theoretical temperature--electric field ($T$--$E$) phase diagram associated with the free-energy given by Eq.~\ref{eq:potential} for $\beta > 0$ and $\delta > 0$. Hatched and hatched-dotted curves represent, respectively, second-order transition and limit of stability curves. The thermodynamic paths for $T>T_c$, $T_c>T>T_0$ and $T<T_0$ are described in the text.}
\label{fig:3}
\end{figure}

Figure~\ref{fig:3} shows the location of the PA, AFE and field-induced FE and FI or WF phases in a theoretical temperature-field $T$--$E$ phase diagram. For $T\ge T_c$ the PA phase ($\eta = 0$, $P = 0$)  stable at $E=0$ transforms into the FE phase ($\eta = 0$, $P = \chi_0E$) for $E\neq 0$. For $T_c>T>T_0$ the AFE phase $[\eta = \pm(-\frac{-\alpha}{\beta})^{\nicefrac{1}{2}},P=0]$ stable at $E=0$, transforms into a FI (or WF) phase for $E\neq 0$, in which the equilibrium values of $\eta$ and $P$ are given by $\eta = \pm [(-\alpha -\delta P^2)/\beta]^{\nicefrac{1}{2}}$ , where $P$ is a real root of the Cardan equation
\begin{equation}
\frac{\delta^2}{\beta}P^3-\left(\frac{1}{\chi_0}-\frac{\alpha\delta}{\beta}\right)P + E = 0
\label{eq:4}
\end{equation}
With increasing field the FI phase transforms across the second-order transition curve
\begin{equation}
E = \pm \frac{1}{\chi_0}\left(-\frac{\alpha}{\delta}\right)^{\nicefrac{1}{2}}
\label{eq:5}
\end{equation}
into the FE phase, which implies that the FI space-group is a subgroup of the FE space-group. For $T<T_0$ the transformation of the FI into the FE phase becomes first-order, crossing the region of coexistence of the FI and FE phase. With increasing field the FI phase transforms discontinuously into the FE phase, the limit of stability of the FI phase corresponding to the curve $E_{c1}$, the equation of which is given by the condition $\partial_{PP}^2\Phi.\partial_{\eta\eta}^2\Phi-\left[\partial_{\eta P}^2\Phi\right]^2 = 0$ with ($\eta\neq 0$, $P\neq 0$) corresponding to:
\begin{equation}
\frac{3\delta^3}{\beta}P^4 + \delta P^2\left(\frac{4\delta\alpha}{\beta}-\frac{1}{\chi_0}\right) + \alpha\left(\frac{\delta\alpha}{\beta}-\frac{1}{\chi_0}\right) = 0
\label{eq:6}
\end{equation}
The $E_{c1}(T)$ curve is obtained by introducing a real root of Eq.~\ref{eq:4} into Eq.~\ref{eq:6}. With decreasing field the FE phase reaches its limit of stability on the $E_{c2}$ curve corresponding to Eq.~\ref{eq:5}. The merging point of the $E_{c1}$ and $E_{c2}$ curves provides the values of $T_0$ and $E_0$. The first-order transition curve between the FI and FE phases, not shown in Fig.~\ref{fig:3}, is located between the $E_{c1}$ and $E_{c2}$ curves below $T_0$. Its equation is given by $\Phi(P_\mathrm{FI},\eta_\mathrm{FI}) = \Phi(P_\mathrm{FE},0)$ where $P_\mathrm{FE} = \chi_0 E$, $P_\mathrm{FI}$ is a real root of Eq.~\ref{eq:4}, and $\eta_\mathrm{FI} = \pm \left[(-\alpha - \delta P^2_\mathrm{FI})/\beta\right]^{\nicefrac{1}{2}}$.

One can deduce from Eq.(\ref{eq:eos2}) the temperature dependence of the dielectric susceptibility  at the PA$\rightarrow$AFE transition. For a second-order transition ($\beta > 0$) one gets below $T_c$:
\begin{equation}
\chi(T) = \displaystyle\frac{\chi_0}{1+\delta a \chi_0\displaystyle\frac{(T_c-T)}{\beta}}     
\label{chiT}
\end{equation}
the temperature dependence of which depends on the sign of $\delta$ (Fig. 4 (a)). Fig. 4 (b) shows $\chi(T)$ for a first-order transition ($\beta < 0$) occurring at $T_1>T_c$, which involves an upward ($\delta < 0$)  or downward ($\delta > 0$) discontinuity. AFE transitions verifiy the preceding temperature dependences of $\chi(T)$ for $\delta > 0$, with a downward discontinuity at $T_1$ for CsH$_3$(SeO$_3$)$_2$ \cite{Makita1965}, Cu(HCOO)$_2$.4H$_2$O \cite{Okada1965}, KCN \cite{Gesi1972,Ortiz-Lopez1988}, NH$_4$H$_2$PO$_4$ \cite{Mason1952} and PbZrO$_3$ \cite{Kanzig1957} and a decrease below $T_c$ for C$_4$O$_4$H$_2$ \cite{Feder1976}.
 
Eq.~(\ref{eq:4}) provides the electric field dependence $P(E)$ of the polarization. It corresponds to a double hysteresis loop which, as shown in Fig.~\ref{fig:3}, can be observed for a second-order AFE transition below a temperature $T_0<T_c$ (Fig.~\ref{fig:hysteresis} (a)--(d)), whereas for a first-order AFE transition it is observed below $T_1>T_c$  (Fig.~\ref{fig:hysteresis} (e)--(f)). Characteristic double AFE loops have been observed in a number of AFE transitions, e.g. in CsH$_3$(SeO$_3$)$_2$ \cite{Makita1965}, Cu(HCOO)$_2$.4H$_2$O \cite{Okada1965} or PbZrO$_3$ \cite{Kanzig1957}. Double hysteresis loops are also reported at first-order FE transitions, as for example in BaTiO$_3$ \cite{Merz1953}. However at variance with AFE transitions the two loops are observed within the region of stability of the PA phase and merge into a single loop at the transition to the FE phase.

\begin{figure}[htbp]
\begin{center}
\end{center}
\includegraphics[width=0.48\textwidth]{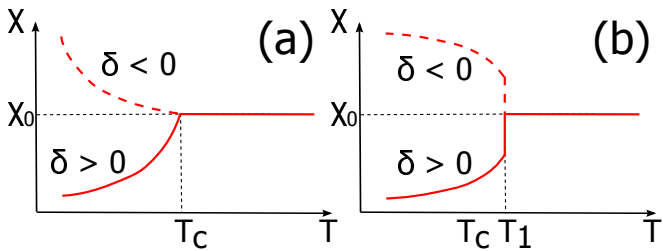}
\caption{Temperature dependence of the dielectric susceptibility $\chi(T)$ given by Eq. (4) across a second-order (a) and first-order (b) transition.}
\label{fig:4}
\end{figure}

\begin{figure}[htbp]
\begin{center}
\includegraphics[width=0.48\textwidth]{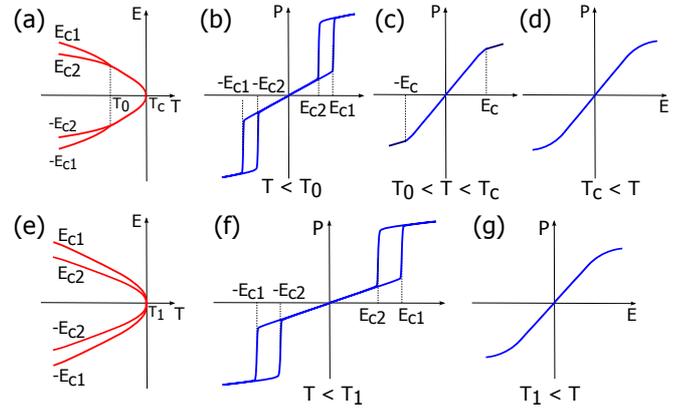}
\end{center}
\caption{$E(P)$ and $E_c(T)$ curves deduced from equations (2) and (3) for a second-order PA--AFE transition (a)--(d) and a first-order transition (e)--(g). At low fields the linear behaviour of $P(E)$ indicates a progressive transformation of the AFE phase into a FI or WF dipolar order. At a coercive field $E_{c1}$ a discontinuous transition occurs to a FE phase. Reversing the field, $P(E)$ decreases along a different path below a coercive field $E_{c2}<E_{c1}$ forming a loop before reaching the linear regime.}
\label{fig:hysteresis}
\end{figure}

A biquadratic $\delta\eta^2 P^2$ coupling always exists at AFE transitions. However, a $\mu\eta P^n$ ($n\ge 2$) coupling can also be permitted by symmetry, which  modifies the previous results. The corresponding equations of state:
\begin{eqnarray}
\eta(\alpha + \beta \eta^2 + \gamma \eta^4 + \delta P^2) + \mu P^n  & = & 0\\
P(\frac{1}{\chi_0} + \delta\eta^2 + n\mu\eta P^{n-2} + \nu P^2) & = & E
\end{eqnarray}
do not allow a stable FE phase under applied field but only FI or WF ($\eta\neq 0$, $P\neq 0$) phases, the stabilization of which requires taking into account an additional $\frac{\nu}{4}P^4$ invariant in $\Phi$. A double hysteresis loop can occur at a first-order field-induced phase transition between two isostructural FI or WF phases having different regions of stability. The temperature dependence of the dielectric susceptibility $\chi(T)$ has a similar shape than for a $\delta\eta^2P^2$ coupling.  

$\eta P^2$ couplings exist exclusively for "proper" ferroelastic transitions \cite{Toledano1980} where $\eta$ has the symmetry of a spontaneous strain. Phase transitions in DyVO$_4$ \cite{Unoki1977,Kishimoto2010}, TeO$_2$ \cite{Peercy1974}, BiVO$_4$ \cite{Bierlein1975} and NdP$_5$O$_{14}$ \cite{Schulz1974} verify this property. As shown in Table~\ref{tab} application of  electric fields to the ferroelastic phases of the four compounds induce a $\eta P^2$ coupling for $E_{xy}$ fields, with the emergence of FI or WF phases, or only a $\eta^2P^2$ coupling for fields along $z$, giving rise to a FE phase. AFE dielectric anomalies have been reported at the transitions in DyVO$_4$ \cite{Unoki1977,Kishimoto2010} and TeO$_2$ \cite{Peercy1974}. $\eta P^3$ couplings are allowed at \emph{ferroelastoelectric} transitions \cite{Toledano1977} where the AFE order-parameter has the symmetry of a third-rank piezoelectric tensor component. $\eta P^4$ couplings are found in proper \emph{ferrobielastic} transitions \cite{Toledano1977} where the order-parameter has the symmetry of an elastic stiffness.

\section{Discussion}
\label{sec:analogy}

Our extended investigation of structural transitions to non-polar phases shows that although conditions 1 and 2 are satisfied in a large number of materials, the emergence of a FE phase above a coercive field may not always occur because of the large energy difference between the low-field FI, and  high-field FE phases. This is due, in particular, to the additional strains possibly required for stabilizing the FE phase. For example, at the AFE transition in NH$_4$H$_2$PO$_4$, a double hysteresis loop could not be observed \cite{Kanzig1957}, the onset of a FE phase implying an orthorhombic deformation of its monoclinic or triclinic FI phase. Therefore, antiferroelectrics presenting all the dielectric features currently assumed for PA--AFE transitions should not constitute a widespread class of materials, as compared to ferroelectrics. This is in contrast to antiferromagnets which form the largest class of magnetically ordered materials. However, our proposed definition of PA--AFE transitions, given in sec.~\ref{sec:symmetry}, does not imply the stabilization of a FE phase under applied field or a double hysteresis loop, and extends the current characterization of antiferroelectrics to the larger class of materials in which a polar (FI or WF) field-induced phase emerges from a non-polar phase.

The analogy between antiferromagnets and antiferroelectrics is usually invoked because both antiferromagnetic and AFE structures display antiparallel arrays of spins or dipoles and because ferromagnetic or FE phases emerge above coercive fields. Our work underlines a deeper analogy consisting of the common microscopic nature of the symmetry-breaking order-parameter in the two classes of materials: the emergence of a discrete array of local dipolar sites, which constitutes in our approach the microscopic symmetry-breaking mechanism for the formation of an AFE phase, is the structural analogue of the continuous microscopic spin-density waves which are the symmetry-breaking order-parameters at antiferromagnetic transitions. It can be questioned if the analogy extends at the level of the interactions between dipoles i.e. if there exists a physical criterion characterizing AFE interactions, similar to the negative sign of the exchange interaction typifying antiferromagnets. In this respect it can be noted that all experimental examples of AFE transitions show a decrease of the dielectric permittivity, corresponding to a positive sign of the coupling coefficient $\delta$ between $\eta$ and $P$. Such repulsive coupling is necessary for compensating the attractive (negative) interactions existing between antiparallel dipoles \cite{Margenau1969} and results from the different types of repulsive forces \cite{London1937} between permanent and induced dipoles. By contrast, the dielectric permittivity at improper ferroelectric transitions undergoes an upward discontinuity \cite{Levanyuk1974} reflecting the attractive coupling between $\eta$ and $P$ required for compensating the repulsive interactions between parallel dipoles. Although such considerations need to be substantiated by a detailed theoretical analysis they suggest that the decrease of the dielectric permittivity at a structural transition to a non-ferroelectric phase denotes the presence of AFE interactions, in the same way that the shape of its magnetic susceptibility typifies an antiferromagnetic ordering. 

The essential difference, noted in the introduction of this article, between the time-reversal symmetry involved in magnetic transitions and the localized symmetries characterizing structural transitions, does not preclude a formal analogy between the theoretical approaches to magnetic and structural transitions, and this analogy can be used for deducing the still unknown fundamental properties of antiferroelectrics from the well established properties of antiferromagnets. However, one should keep in mind that although spin densities are localized, electric dipole densities localized at polarisable sites represent only a conceptual image which has been used for establishing the dielectric properties of polarized insulators. In the case of AFE transitions this image can be specified in two different ways depending on the symmetry breaking mechanism occurring at the transition. If the transition induces a breaking of the translational periodicity associated with a wave-vector $k\neq 0$, the AFE structure can be considered as formed by an alternation of unit-cells having antiparallel local polarizations. For transitions without modification of the crystal unit-cell ($k=0$) one can separate in each unit-cell two regions corresponding to antiparallel polarizations. In both cases the dipoles are assumed to be located at sites consistent with the symmetry operations of the AFE structures and represent the polarization of the entire surrounding volumes (unit-cell or region of a unit-cell).

The relation of the AFE order-parameter $\eta$ with the local distribution of dipoles can correspond to two different transition mechanisms, similar to the different mechanisms characterizing "proper" and "improper" ferroelectric transitions \cite{Levanyuk1974,Toledano1976}. For "proper" AFE transitions $\eta$ can be directly expressed in terms of the local dipole distribution either in a continuous formalism, as a polarization wave amplitude, or in a discrete formalism, as a linear combination $\Sigma (\vec p_i - \vec p_j)$ of local dipoles belonging to antiparallel arrays of emerging polar sites. For "improper" AFE transitions $\eta$ represents a structural (displacive or ordering) mechanism  which typifies the lowering of symmetry at the transition, the emergence of an antiparallel polarization wave amplitude being an induced secondary effect of the preceding primary mechanism.

\section{Summary and conclusion}
\label{sec:summary}

In summary, AFE phase transitions have been shown to occur under combined local and macroscopic symmetry conditions, which provide a symmetry based definition of this class of structural transitions. A Landau theoretical description of their dielectric properties has been given by taking into account the electric-field induced couplings existing between the AFE and polarization order-parameters. This description leads to properties of AFE materials differing essentially from the properties deduced from Kittel's model of antiferroelectrics \cite{Kittel1951}: the FE phase is absent from the phase diagram at zero field, its emergence as a purely field induced effect being conditioned by symmetry requirements. Furthermore, the AFE order parameter $\eta$ may represent a structural mechanism inducing indirectly the antiparallel dipole lattices ("improper" antiferroelectricity), or can be expressed directly in terms of antiparallel dipoles ("proper" antiferroelectricity) as assumed by Kittel. 

The authors are grateful to B. Mettout for very helpful discussions and J. Kreisel for suggesting this work. Work supported by the National Research Fund, Luxembourg (FNR/P12/4853155/Kreisel).


\end{document}